\documentclass[aip,rsi,reprint]{revtex4-1}

\usepackage{amsmath}
\usepackage{amssymb}

\usepackage{graphicx}
\usepackage{dcolumn}
\usepackage{bm}

\usepackage[utf8]{inputenc}
\usepackage[T1]{fontenc}
\usepackage{mathptmx}
\usepackage{etoolbox}

\makeatletter
\def\@email#1#2{%
 \endgroup
 \patchcmd{\titleblock@produce}
  {\frontmatter@RRAPformat}
  {\frontmatter@RRAPformat{\produce@RRAP{*#1\href{mailto:#2}{#2}}}\frontmatter@RRAPformat}
  {}{}
}%
\makeatother
\begin{document}

\preprint{AIP/123-QED}

\title[Rapid, Broadband, Optical Spectroscopy of Cold Radicals]{Rapid, Broadband, Optical Spectroscopy of Cold Radicals}
\author{Ashay N. Patel}
\author{Madison I. Howard}
\affiliation{ 
Division of Physics, Mathematics, and Astronomy,\\ California Institute of Technology, Pasadena CA 91125, USA}
\author{Timothy C. Steimle}
\affiliation{ 
Division of Physics, Mathematics, and Astronomy,\\ California Institute of Technology, Pasadena CA 91125, USA}
\affiliation{School of Molecular Science, Arizona State University, Tempe AZ 85287, USA}\author{Nicholas R. Hutzler$^*$}
 \email{hutzler@caltech.edu}
\affiliation{ 
Division of Physics, Mathematics, and Astronomy,\\ California Institute of Technology, Pasadena CA 91125, USA}

\date{\today}

\begin{abstract}
Optical spectroscopy of molecular radicals is an important tool in physical chemistry, and is a prerequisite for many experiments which use molecules for quantum science and precision measurement.  However, even the simplest molecules have complex spectra which can be very time consuming to measure.  Here we present an approach which offers the ability to measure the optical spectra of cryogenically-cooled molecular radicals with much greater efficiency.   By combining a supercontinuum laser with a cryogenic buffer gas molecular source and a commercial optical spectrometer, we realize 15~nm of simultaneous bandwidth with 0.56~pm $(\approx 0.5$~GHz) resolution and high sensitivity.  As a demonstration we measure and assign hundreds of lines and dozens of molecular constants from 15 bands in the $B^2\Sigma^+-X^2\Sigma^+$ system of CaF, including a low-abundance isotopologue, in a few hours.  The setup is robust, simple, and should enable spectroscopy of molecular radicals with much higher throughput.
\end{abstract}

\maketitle

\section{\label{sec:intro} Introduction}

Spectroscopy is foundational in both physics and chemistry, and has informed much of what we know about the physical universe. Optical spectroscopy of gas-phase molecules is particularly informative as it probes individual quantum levels including their response to the physical environment~\cite{Herzberg1966Polyatomic}, making it determinative for experiments using molecules for precision measurements of fundamental physics~\cite{Hutzler2020PolyReview,Safronova2018Review} and quantum information science~\cite{Cornish2024Review}.  However, obtaining spectra with sufficient resolution is often a significant burden; it is common for experiments in these areas to spend years obtaining the data needed to understand the underlying quantum structure of a molecule.  Modern quantum science and precision measurement experiments present particular challenges.  First, these experiments typically use low-lying rotational levels, requiring low temperature ($\lesssim$100~K) methods to sufficiently populate those states.  Second, these species are nearly always reactive, refractory, or both; continuous sources such as ovens are impractical in light of the previous point, so these samples are necessarily transient and relatively low density.  Third, these experiments often need up to a dozen (or more) single quantum state resolution vibronic transitions for state control, detection, laser cooling, repumping, etc., often spanning hundreds of nanometers~\cite{Fitch2021Review,Augenbraun2023PolyLCReview}.  Fourth, these experiments typically require high resolution on myriad hyperfine and rotational levels within each vibronic transition.  In combination, these requirements result in a significant time investment, which slows the rate of progress in cold molecule science.

In this manuscript we report a method for rapid, broadband, and high-resolution spectroscopy of radical molecules created in a pulsed cryogenic buffer gas cell (CBGC)~\cite{Hutzler2012}.  The method shows 0.56~pm ($\approx 560$~MHz) resolution with 15~nm range in a single pulse of molecules created from ablation, with near-shot-noise limited absorption sensitivity of $\sim$0.01/pulse.  The approach requires a commercially-available light source and custom spectrometer based on a virtually imaged phased array (VIPA).  We demonstrate this method by studying multiple bands and isotopologues of the $B^2\Sigma^+-X^2\Sigma^+$ system of CaF around 531~nm by ablating a refractory precursor.  We map out the (0,0) vibrational band in around one second with a single repetition of the experiment. By averaging $\sim10,000$ repetitions taken in approximately three hours we resolve up to the (16,16) vibrational band, determine dozens of molecular constants with $\sim$50 MHz precision from hundreds of lines, and observe a low-abundance calcium isotopologue.

There are alternative approaches to broadband, high-resolution spectroscopy, though ours offers a number of advantages tailored to the needs of current quantum science and precision measurement applications.  Fourier Transform infrared (FTIR) spectroscopy methods are challenging to implement in short-lived samples of radicals.  Dispersed laser-induced fluorescence (DLIF)~\cite{Nguyen2018SrOHBR,Zhang2021BR} gives broadband spectral information with rotational resolution, but still requires scanning a light source with a narrow bandwidth.  Frequency combs combined with VIPAs give high resolution and many spectral channels, but are typically used in the IR and require complex optical setups~\cite{Adler2010, Nugent-Glandorf2012VIPA}.  Here we offer an approach which is simple, relatively inexpensive, and tailored to radical, refractory, or otherwise challenging species.

\section{\label{sec:apparatus} Method Overview}

\begin{figure}
\centering
\includegraphics[width=0.7\linewidth]{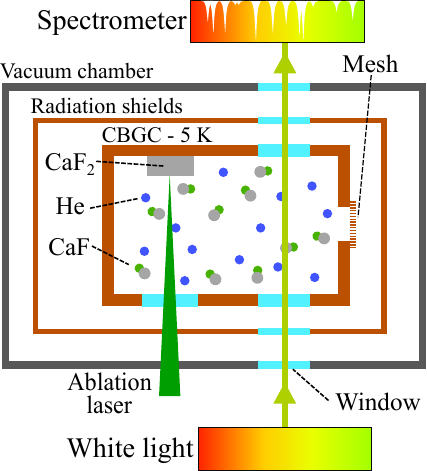}
\caption{Schematic overview of the experimental setup.  Molecules are produced in a cryogenic buffer gas cell (CBGC) via laser ablation of a solid target.  A supercontinuum laser is sent through the molecular cloud and then dispersed on the optical spectrometer.}
\label{fig:apparatus}
\end{figure}

The method is a variation of white-light absorption spectroscopy: white light is passed through a molecular cloud, and comparing the transmitted spectrum with and without the molecules present yields the molecular absorption spectrum.  A supercontinuum laser generates $\sim$1~W of white light, covering the range of 500--2,000~nm.  The light passes through a cloud of cryogenically-cooled atoms or molecules inside a CBGC, then enters a custom spectrometer manufactured by LightMachinery Inc. which combines a  virtually imaged phased array (VIPA) and diffraction grating to disperse and image fiber-coupled light onto a large-area sCMOS camera.  The important design features, which will be discussed in detail below, are the use of an optical spectrometer with resolution comparable to the Doppler width of typical atoms and molecules of interest at a temperature of $\sim$5~K, the high atomic and molecular densities offered by the CBGC method, and a choice of laser and analysis routine to realize near-shot-noise-limited sensitivity.  The integrated spectrometer system is shown schematically in Fig. \ref{fig:apparatus}.

\subsection{\label{subsec:cbgb} Cryogenic Buffer Gas Molecular Source}

We use a cryogenic buffer gas cell (CBGC) to create samples of gas-phase atoms and molecules at 5~K using a helium buffer gas~\cite{Hutzler2012}.  The source has been used in previous experimental studies and is described elsewhere~\cite{Zeng2023Cycling}.  In summary, a pulse tube refrigerator cools a copper cell with approximately 2 inch cubic geometry.  There is a 1.5 inch diameter hole bored through the cell to create an internal volume, with smaller perpendicular holes bored through for installation of ablation targets and optical access for the white light.  A copper tube flows helium into the cell, which exits through a 6~mm diameter aperture.  A 100-mesh copper sheet was epoxied to cover the aperture in order to increase the density of molecules near the absorption probe path when helium is flowed. 

Molecules are produced via laser ablation of targets with $\sim$50~mJ, 6 ns pulses generated from a 532~nm Nd:YAG laser.  For the CaF molecular data presented in this manuscript, our ablation target was a CaF$_2$ sputtering target\footnote{99.9$\%$ purity, 1/8 inch thick, supplied by Heeger Materials.}.  The produced molecules rapidly thermalize translationally and rotationally with 5~K helium atoms. Targets are ablated  with a helium buffer gas flow rate of 20 standard cubic centimeters per minute.  The cell temperature is typically between 5.1 and 5.2~K while ablating at a typical repetition rate of 2~Hz.  The resulting pulse of molecules lasts for a few ms before the molecules either flow out of the cell or diffuse to the walls.

\subsection{\label{subsec:HF spec} HyperFine Spectrometer with Sub-pm Resolution}

The spectrometer used in this manuscript was a custom-designed and manufactured HyperFine spectrometer from LightMachinery Inc.  The spectrometer combines a high-efficiency and high-resolution VIPA etalon that provides large angular dispersion for light of different wavelengths. An orthogonally aligned diffraction grating separates degenerate VIPA mode orders. The resulting 2D pattern of stripes is imaged on a large area sCMOS camera. Included software converts the camera image into a 1D spectrum.  The spectrometer is quoted by the manufacturer to have 0.55~pm resolution with a range from 525~nm and 540~nm; for the data presented here, the instrument was sampled at 0.1~pm.  The spectrometer design is currently being modified by LightMachinery to incorporate the ability to tune the 15~nm window over a much wider range. Coupling into the spectrometer is achieved with a single mode optical fiber.  The optical throughput and camera quantum efficiency give an overall detection efficiency estimated by the manufacturer to be $\sim10\%$.

\subsection{\label{subsec:whitelight} Broadband Light Source}

Maximum sensitivity will be achieved when the spectrometer is limited by photon shot noise, which requires near-saturation of the detector during the $\sim$1~ms long molecular pulse.   As discussed in section \ref{sec:noise}, we would like to detect around 400,000 camera photoelectrons per spectral element (se), which is around 0.5~pm wide.  For a beam with $\sim 1$~mm$^2$ area and  $\lambda\approx$532~nm, this means that we require a spectral irradiance $F$ of
\begin{eqnarray*}
    F & = &  \frac{(400,000~e^-/\mathrm{se})(hc/\lambda)}{(0.1~e^-/\mathrm{photon})(0.5~\mathrm{pm/se})(1~\mathrm{ms})(1~\mathrm{mm}^2)},\\
    & \approx & 3~\mu\mathrm{W/(nm\times mm^2)}.
\end{eqnarray*}
Importantly, this light must be coupled into a single-mode fiber, and there will be losses from reflections, etc. so a light source with $>10~\mu$W/nm spectral intensity is desirable.  Our light source is a supercontinuum laser\footnote{NKT SuperK EVO ERL-04} with $\sim$0.3~mW/nm spectral intensity out of a single mode fiber, which we can couple with reasonable efficiency back into a single mode fiber using standard collimating lenses and fibers since we are only using a limited $\sim$15~nm wide window at a given time.  The output of the supercontinuum laser is sent through a tunable filter\footnote{NKT SuperK VARIA} to dump unused wavelengths for safety reasons.

\subsection{\label{subsec:integration} System Operation}

An electronic trigger pulse from a computer control system prepares the spectrometer to record a spectrum, and after a time delay another trigger pulse fires the Nd:YAG to ablate the target.  An optical shutter\footnote{SRS SR470} allows light from the supercontinuum laser to pass through the cloud for approximately 4~ms, which is the duration of the molecular pulse where the absorption signal is roughly maximized.  The spectrometer camera exposure time is much longer, typically $\sim$100~ms, due to the finite frame rate; however, since the camera has negligible dark count rates, this additional exposure time does not impact the sensitivity.  A second exposure with similar timing occurs after approximately 500~ms, though without the ablation laser firing.  This second exposure gives a background measurement so that an absorption fraction can be calculated.  Note that the absorption laser only takes a single pass through the molecular cloud, which has an absorption length of approximately 1~inch; increasing this with a longer cell or using multi-pass methods would increase the absorption signal, but a single pass is sufficient for our current study.

\subsection{Extraction of Molecular Spectra}

Here we describe how molecular spectra are extracted from the data. A single shuttered exposure gives a single spectrum of counts vs. discrete frequency values $c(f_j)$, both of which are extracted from the camera image after applying the unwrapping algorithm provided with the spectrometer software.  For each pulse of molecules from the source, indexed by $i$, we save two spectra, $c^s_i(f_j)$ and $c^b_i(f_j)$, the signal ($s$) spectrum with the molecules present and the background ($b$) spectrum with no molecules present, the latter of which is obtained by taking a spectrum without firing the Nd:YAG laser.  From this pair of points we compute an uncorrected absorption spectrum
\begin{equation}
    a'_i(f_j) = \frac{c^b_i(f_j)-c^s_i(f_j)}{c^b_i(f_j)}.
\end{equation}

\begin{figure*}
    \centering
    \includegraphics[width=\linewidth]{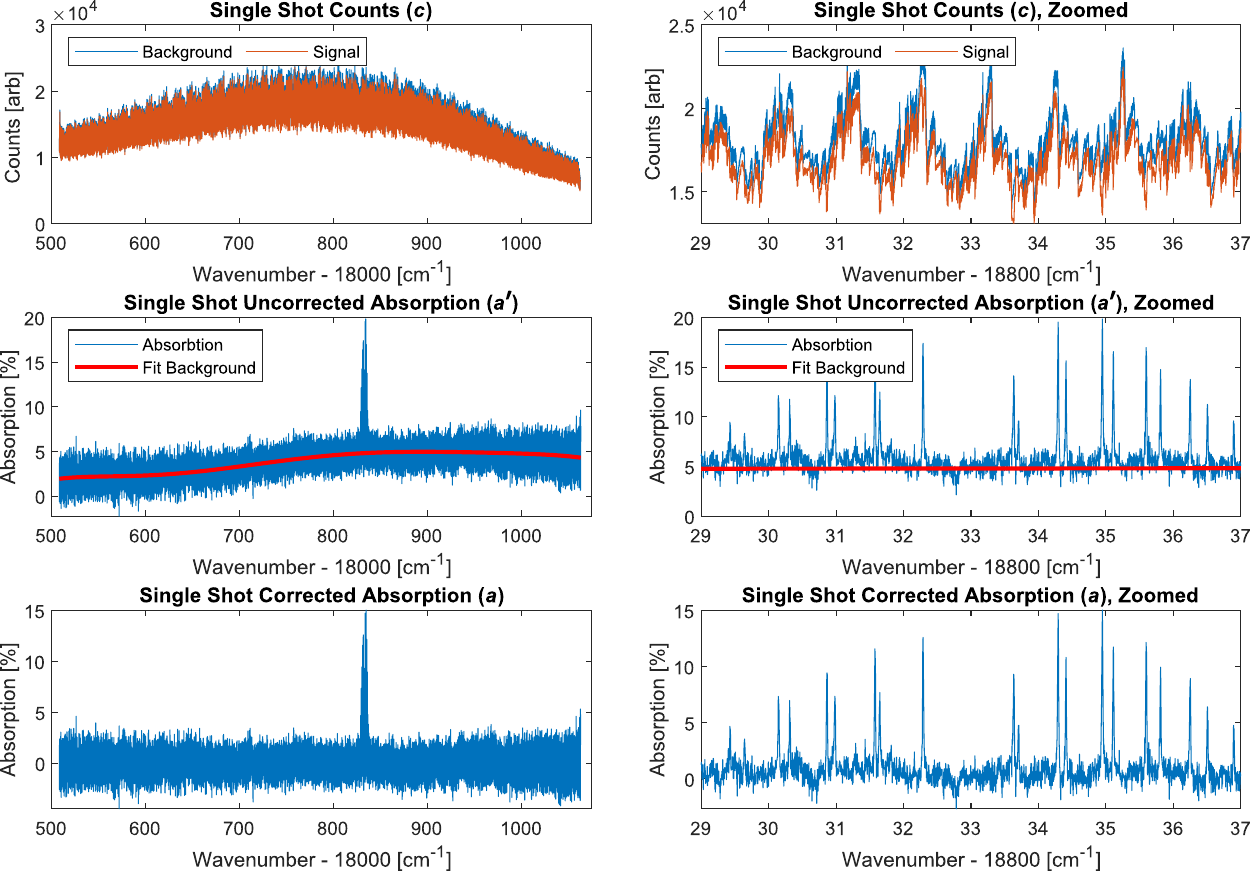}
    \caption{The procedure for going from counts to absorption.  All plots on the right are zoomed-in versions of the plots on the left. 
 \textit{Top}: Counts vs. spectral element for single shots with and without molecules present.  Counts is as reported by the LightMachinery software and is the average photoelectron count over the pixels combined in each spectral bin.  We can see the slowly-varying envelope from the light source and the oscillations on the $\sim$1~cm$^{-1}$ scale due to etaloning.  \textit{Middle}: Uncorrected absorption signal, including polynomial fit to the background.  \textit{Bottom}: Corrected absorption signal after subtracting the polynomial background.}
    \label{fig:singleShotCounts}
\end{figure*}

Due to a frequency-dependent drift of the white light source coupled through the apparatus and into the spectrometer, $a'$ has a fairly large but slowly varying (compared to the instrument resolution) offset; improved alignment could reduce this offset, and could also eliminate the need to save a background shot paired with every signal shot.  We fit this ``slow background'' offset for each pulse as a sextic polynomial $p_i(f_j)$.  We then compute the corrected absorption spectrum
\begin{equation}
    a_i(f_j) = a'_i(f_j) - p_i(f_j).
\end{equation}
For large enough signals, a single corrected absorption spectrum $a$ can be used to extract spectral information, assuming that the unit has been calibrated (see section \ref{sec:i2cal}).  A set of typical $a',p,$ and $a$ for CaF are shown in Fig.~\ref{fig:singleShotCounts}.

The signal-to-noise ratio (SNR) of the spectrum can be improved by averaging.  However, before averaging the spectra together we must take into account that the frequency calibration of the instrument drifts in time.  For spectra with a sufficiently high SNR on a single shot, as is the case for CaF, we can identify known lines and shift the frequencies so that the peaks occur at the same position.  For the CaF data presented here, we calibrate using the known~\cite{Devlin2015} $^PQ_{12}(1)$ line at 18832.445~cm$^{-1}$ and $P_1(1)$ line at 18832.441~cm$^{-1}$, which are overlapped within the instrument resolution.  We find that the frequency calibration drifts on the order of 0.01~cm$^{-1}$/hour.  For spectra with insufficient SNR, a known calibration species could be produced simultaneously, or fluorescence from a calibrated lamp could be overlapped with the white light to provide a known line position.  After shifting the frequencies so that the single-shot lines overlap, we can form the averaged corrected absorption spectrum
\begin{equation}
    \bar{a}(f_j) = \sum_i a_i(f_j)/N,
\end{equation}
where $N$ is the number of absorption spectra to be averaged together.

\begin{figure}
    \centering
    \includegraphics[width=\linewidth]{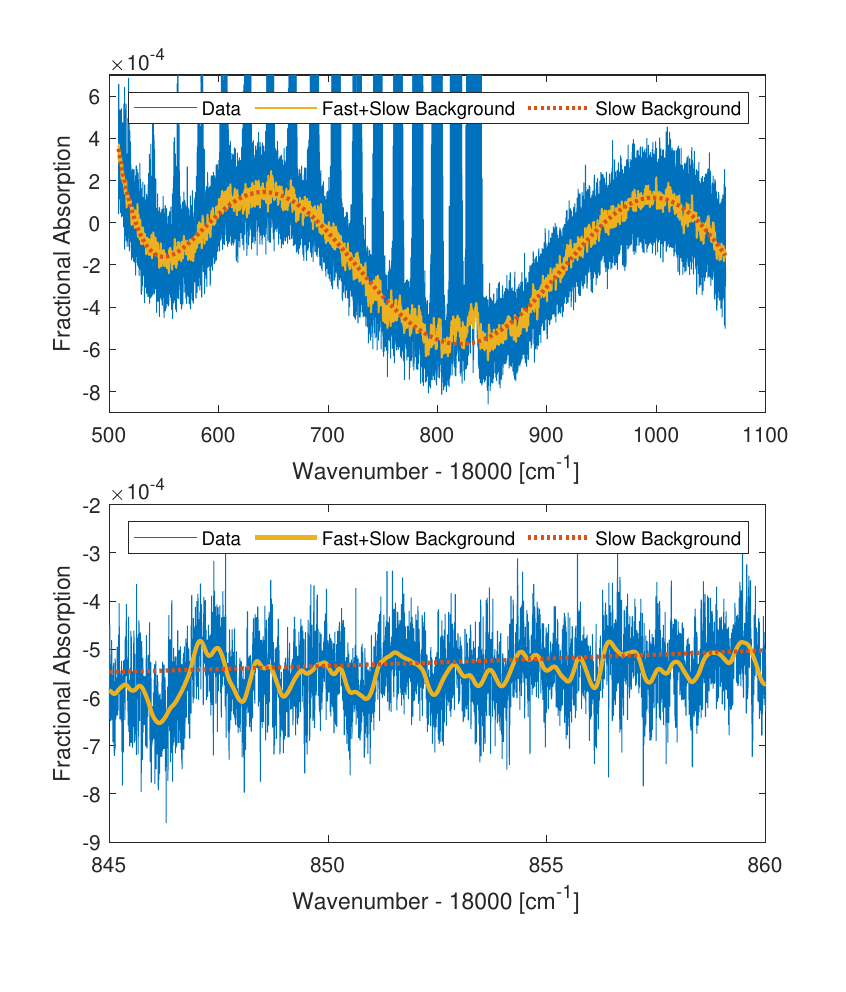}
    \caption{Background subtraction for spectra.  Top: entire spectrum, showing the uncorrected, averaged absorption spectra along with the slow background and combined (slow and fast) background.  Note that near the strong $(0,0)$ band head the fitted fast background appears to be influenced by the spectra lines, but on a level of $\sim 10^{-4}$, which is negligible compared to the absorption feature size of $\sim 10^{-1}$.  This is an artifact due to the high density of strong spectral lines in this region.
 Bottom: zoom on a region free of molecular signal showing the different behaviors of the slow and fast backgrounds.}
    \label{fig:background}
\end{figure}

\begin{figure}
    \centering
    \includegraphics[width=\linewidth]{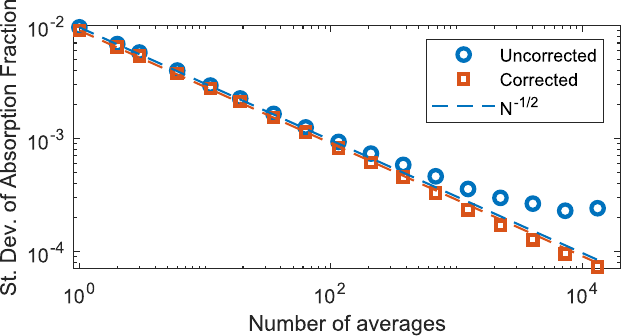}
    \caption{Noise vs. number of averages $N$.  The noise is reduced as $\sim N^{-1/2}$.  Below $\sim 2\times10^{-4}$ the contribution from spectral noise begins to contribute, showing a leveling-off of the noise level for the uncorrected spectrum.  The corrected spectrum subtracts this spectral noise so the averaging continues to improve sensitivity.}
    \label{fig:noiseAveraging}
\end{figure}

The averaged corrected spectrum $\bar{a}$ still has some residual spectral noise from the light source and apparatus etaloning on the level of $10^{-4}$ fractional absorption, which begins to appear after $\sim 10^3$ averages.  Since this noise varies slowly in frequency compared to the instrument resolution and the width of Doppler-broadened lines at cryogenic temperatures, it does not complicate spectral assignments.  Nevertheless, we remove the ``slow background" with an eighth-order polynomial fit, and the ``fast background'' noise by fitting a robust spline~\cite{Garcia2010Robust,Garcia2011Robust} set to ignore points above a threshold absorption value of $2\times 10^{-4}$ (after removing the slow background).  These backgrounds are shown in Fig.~\ref{fig:background}.  By including the robust spline fit, we are able to continue averaging down as $N^{-1/2}$ to at least the level of $7.3\times10^{-5}$.  The noise level with increased averaging is shown in Fig.~\ref{fig:noiseAveraging}.

\begin{figure}
    \centering
    \includegraphics[width=\linewidth]{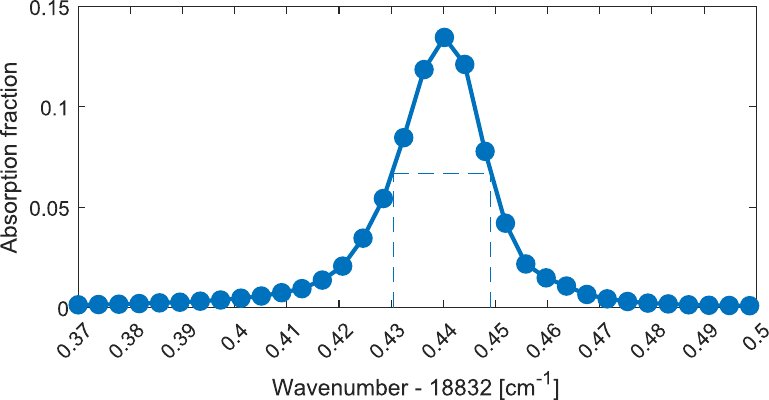}
    \caption{A single observed line showing the experimental resolution of 0.019~cm$^{-1}$.  This feature arises from the overlapped $^PQ_{12}(1)$ and $P_1(1)$ lines in the $(0,0)B^2\Sigma^+-X^2\Sigma^+$ band of CaF, which differ by 0.004~cm$^{-1}$ and are therefore not resolved.  The instrument is sampled at 0.1~pm.}
    \label{fig:singleLine}
\end{figure}

To determine the resolution, we can observe the $^PQ_{12}(1)$ and $P_1(1)$ lines in the $(0,0)B^2\Sigma^+-X^2\Sigma^+$ band of CaF, which differ by 0.004~cm$^{-1}$ and are therefore not resolved.  We find the full width at half maximum (FWHM) to be 0.019~cm$^{-1}$ or 0.56~pm, in good agreement with the quoted resolution of 0.55~pm, as shown in Fig.~\ref{fig:singleLine}.  Note that the instrument is sampled at 0.1 pm.

\subsection{Iodine Calibration}\label{sec:i2cal}

\begin{figure}
    \centering
    \includegraphics[width=\linewidth]{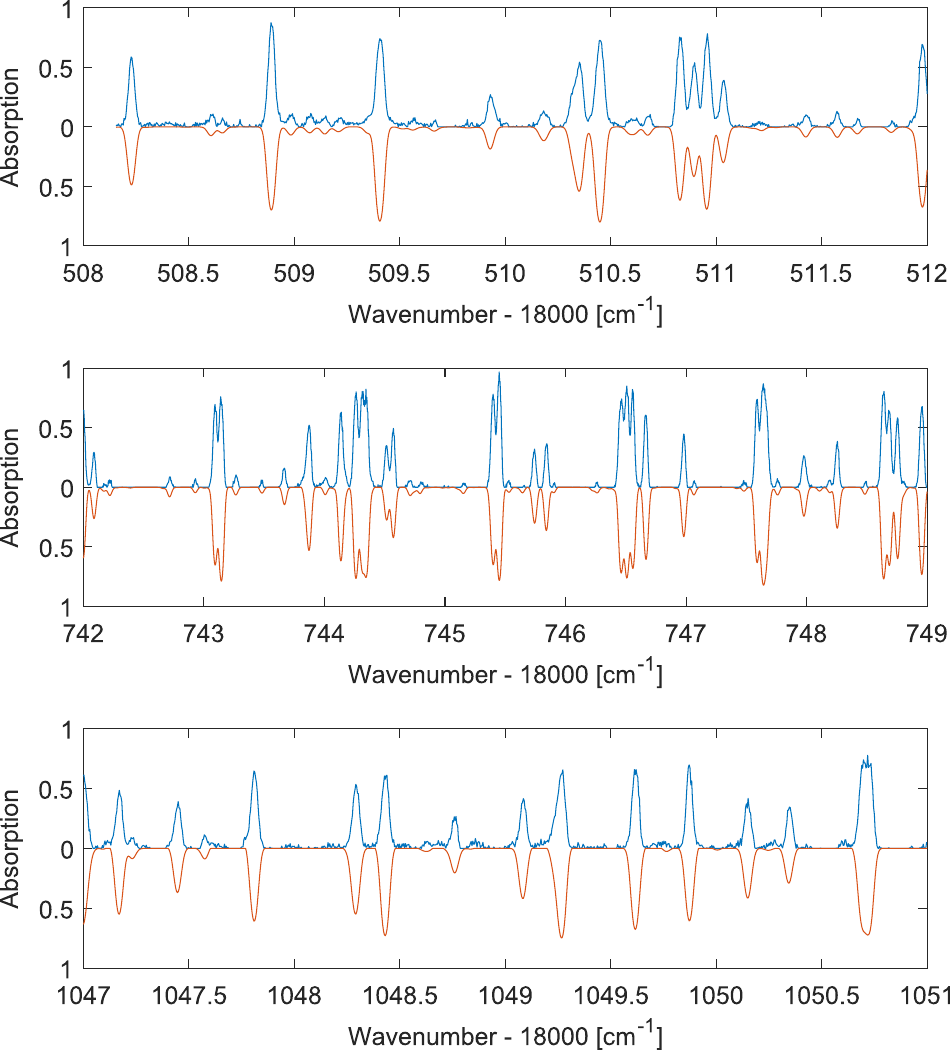}
    \caption{Single-shot calibration with molecular iodine showing the low (top), middle (middle), and high (bottom) wavenumber portions of the spectrometer's range.  On each plot, the upper curve shows the data and the lower curve shows the IOSpec5 prediction.  The prediction has an overall height scaling factor.}
    \label{fig:I2Calibration}
\end{figure}

We record an absorption spectrum of an iodine vapor cell in order to provide an absolute wavelength calibration.  Since we are unable to turn the absorption signal off without modifying the optical path, we instead rely on the fact that the vapor cell absorption signals are rather strong and do not require averaging, and that the absorption features are narrower in frequency than the spectral noise.  To extract the iodine spectrum, we split the optical path using a 50/50 non-polarizing beam splitter, directing one path through the iodine vapor cell and the other around it. We first record a spectrum of counts vs. frequency $c_{cal}^s(f_j)$ with the white light passing through the vapor cell, which is at room temperature and has a single-pass optical path length of $\sim$15~cm, while blocking the other path.  Next, we block the iodine path and record a spectrum $c_{cal}^b(f_j)$ without the iodine vapor in the path. Since this spectrum involves changing the optical path, there is a rather large but slowly-varying systematic shift in the counts versus wavelength.  To correct for this, we compute a smoothed function $s = \langle c_{cal}^s(f_j)/c_{cal}^b(f_j)\rangle$ using a third order Savitzky-Golay filter and then scale $c_{cal}^b(f_j)$ by this function so that the slowly-varying intensity envelopes match.  We then compute the iodine absorption
\begin{equation}
    a_{cal}(f_j) = \frac{s\times c^b_{cal}(f_j)-c^s_{cal}(f_j)}{(s\times c^b_{cal}(f_j)+c^s_{cal}(f_j))/2},
\end{equation}
and then compare this spectrum to one predicted by IOSpec5~\cite{Salumbides2006IOSpec}.  A comparison is shown in Fig.~\ref{fig:I2Calibration}.  The fit agrees to well within the instrument resolution across the entire range.

Taking the IOSpec5 prediction as the true spectrum, we apply a linear transformation to the instrument-reported wavelengths $\tilde{\nu}_\mathrm{inst}$ to determine the calibrated wavelengths $\tilde{\nu}_\mathrm{cal} = m\tilde{\nu}_\mathrm{inst}+b$.  The linear shift is calculated by numerically maximizing the overlap of the true spectrum with the measured spectrum after linear frequency scaling.  In Table~\ref{tab:cal} we report the calculated frequency shift at the red (blue) ends of the spectrum, $\Delta\tilde{\nu}_\mathrm{red}~ (\Delta\tilde{\nu}_\mathrm{blue})$ along with their difference and sum, for four calibration runs spaced over several hours.  We can see that the instrument has an offset of $(-0.164\pm0.001)$~cm$^{-1}$, where the drift is likely due to temperature variations, which the manufacturer quoted as resulting in a shift of around $0.12$~cm$^{-1}$/$^\circ$C.  The linearity $m$ appears to be stable within our ability to measure it over the course of several hours.

As a further check, we can compare against the previously-measured~\cite{Devlin2015} $^PQ_{12}(1)$ line at 18832.445~cm$^{-1}$ and $P_1(1)$ line at 18832.441~cm$^{-1}$ in the $(0,0)B^2\Sigma^+-X^2\Sigma^+$ band.  These lines are overlapped within the instrument resolution.  As shown in Table~\ref{tab:cal}  we measure this feature over the course of several hours to be in the range of $(18832.442\pm0.001)$~cm$^{-1}$, in good agreement.

\begin{table}
    \centering
    \begin{tabular}{ccccc}
    \toprule
$\Delta\tilde{\nu}_\mathrm{red}$ & $\Delta\tilde{\nu}_\mathrm{blue}$  & $m$ & $b$ & Cal. Line \\
cm$^{-1}$ & cm$^{-1}$ & --- & cm$^{-1}$ & cm$^{-1}$ \\
\hline
--0.2080 &  --0.1189 &  1.0002 &  --0.1635 & 18832.443 \\
--0.2073 &  --0.1180 &  1.0002 &  --0.1627 & 18832.441 \\
--0.2094 &  --0.1214 &  1.0002 &  --0.1654 & 18832.442 \\
--0.2069 &  --0.1176 &  1.0002 &  --0.1623 & 18832.442 \\
\toprule
    \end{tabular}
    \caption{The results of four iodine calibration runs spaced over several hours.  }
    \label{tab:cal}
\end{table}

\subsection{Noise}\label{sec:noise}

The absorption fraction noise floor for a single shot is 0.0090, and is dominated by photon shot noise and spectral noise due to etaloning.  Since the spectrometer records light intensity at multiple wavelengths simultaneously, and molecular signals are narrow features in the spectrum, the system is highly robust against laser intensity noise and slow drifts in the spectral properties of the laser.

\begin{figure}
    \centering
    \includegraphics[width=\linewidth]{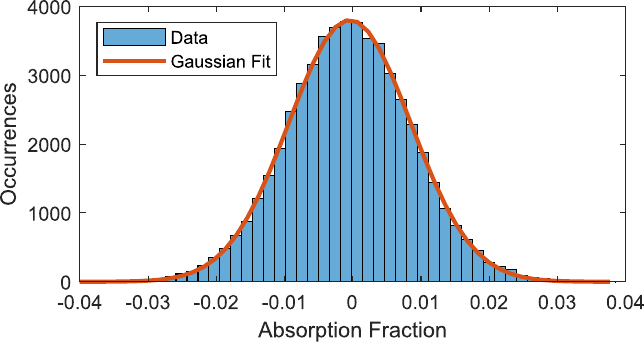}
    \caption{Histogram of absorption fraction occurrences for a single-shot corrected spectrum.  Only measurements above 18,845~cm$^{-1}$ are included since there are no observed lines in that region.  The Gaussian fit has a standard deviation of 0.0089, whereas the data has a standard deviation of 0.0092.}
    \label{fig:noiseHistogram}
\end{figure}

To compute the noise level we look at regions of the spectrum where we do not observe or expect any lines, or alternatively by computing absorption spectra using only the background shots with no molecules present.  Here we define the noise as the standard deviation of the corrected absorption spectrum with no molecules present, which we find is Gaussian (see Fig.~\ref{fig:noiseHistogram}).

We can compare this noise level to the estimated photon shot noise contribution.  The spectrometer camera well depth is 10,000~$e^-$/pixel, so we operate with $\sim$8,000~$e^-$/pixel to avoid saturation.  According to the manufacturer, each spectral element (se) at resolution is obtained by averaging around 50 pixels; we sample the instrument at one fifth of the resolution, so there are around 80,000 $e^-$/se.  The shot noise on this number of photoelectrons is $\sqrt{80,000}\approx 280$.  Per the camera manufacturer the read noise per pixel is typically 2~$e^-$/pixel, with $<1$~$e^-$/pixel of dark counts for our exposure length, so we expect approximately 300 $e^-$/se of total noise.  We estimate the shot-noise limited, single-shot absorption noise level to therefore be $300/80,000 \approx 0.0038$, reasonably close to the measured single-shot value of 0.0090.  Two steps which could get the single-shot value closer to this minimum level are flattening the optical spectrum and reducing etaloning (see Fig.~\ref{fig:singleShotCounts}) to get all spectral elements closer to saturation, reducing drift in the measured spectrum via more rigid optical mounting, and using a camera with a larger well depth.

We find that averaging is very effective at reducing the noise.  With 13,198 averages the noise level is reduced to $7.3\times 10^{-5}$, consistent with a reduction of $\sim (13,198)^{-1/2}$ from the single-shot value.  Further investigation is required to understand the ultimate noise floor of the system.

\subsection{\label{subsec:CaF} Spectroscopy of CaF}

The properties of calcium monofluoride, CaF, are of interest for quantum simulation and quantum computing applications in large part due to its ability to cycle many photons with only small losses to internal states~\cite{Fitch2021Review}.  This enables laser cooling and confinement in optical traps, including tweezer traps, affording single-molecule control~\cite{Anderegg2019Tweezer,Bao2024RSC,Bao2023CaFEntanglement,Holland2023CaFEntanglement}. The initial stage of the cooling and trapping of CaF is the generation of a cold and slow intense molecular beam by skimming the output of a pulsed CBGC. In previous studies, CaF was produced by laser ablation of a solid Ca target in the presence of gaseous SF$_6$ which flows into the CBGC source~\cite{Anderegg2017RF,Truppe2017Slow}. An alternative approach for the generation of CaF in a CBGC source, which is used here, is the ablation of a solid CaF$_2$ target. Here we will show that the new spectrometer has the sensitivity and resolution for CaF internal state characterization.

An additional motivation for developing a broadly tunable spectroscopic technique with sufficient spectral resolution and sensitivity to observe rotational fine structure in vibronic transitions of CaF is as a possible route for monitoring the relative isotopic abundance of Ca. The relative calcium isotope compositions (e.g. $^{44}$Ca/$^{42}$Ca and $^{44}$Ca/$^{43}$Ca) of a sample, which are traditionally measured via multi-collector inductively coupled plasma mass spectrometry (MC-ICP-MS), are particularly relevant to understanding biological~\cite{CostasRodriguez2016} and geological~\cite{Antonelli2021} processes. A simple optical spectroscopic technique for the determination of relative isotopic abundances, such as the one described here, is an attractive alternative to mass spectroscopic methods. The isotopic shifts in Ca atomic transitions are difficult to resolve using optical spectroscopy whereas the rotational and vibrational isotopic spectral shifts in the electronic spectrum of CaF are significantly larger. Recently it was demonstrated that the intensities of the (1,0)$A^2\Pi$-$X^2\Sigma^+$ bandheads of $^{40}$CaF and $^{44}$CaF are readily resolved and can be used to determine the relative $^{40}$Ca and $^{44}$Ca isotopic abundances of a urine sample~\cite{Zanatta2019}.

There is an extensive spectroscopic database for CaF. A nearly complete list of relevant studies can be found in the recently published~\cite{Lavy2025} analysis of the high-temperature emission spectra of the $A^2\Pi-X^2\Sigma^+$ and $B^2\Sigma^+-X^2\Sigma^+$ electronic transitions. In that work, a high-temperature (2500 K) CaF sample was produced using a hollow cathode discharge source and the emission spectrum recorded at a spectral resolution of 0.05 cm$^{-1}$. 

\begin{figure}
    \centering
    \includegraphics[width=\linewidth]{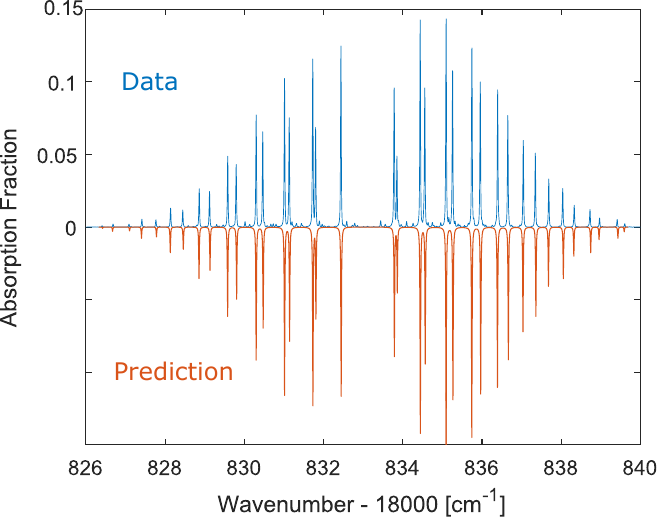}
    \caption{The observed and predicted spectrum for the $(0,0)B^2\Sigma^+-X^2\Sigma^+$ band.}
    \label{fig:CaF00Fit}
\end{figure}

As a diagnostic of the reliability of the spectrometer, the spectral features for the $v=0$ through $v=14$ bands of the $B^2\Sigma^+-X^2\Sigma^+$ electronic transition were measured and used to determine spectroscopic parameters. The 386 measured and assigned spectral features for these fifteen bands are given in Table~\ref{tab:const} of the Appendix.  The observed and predicted spectra for the $(0,0)B^2\Sigma^+-X^2\Sigma^+$ band are shown in Fig.~\ref{fig:CaF00Fit}.  The predicted spectrum was obtained using the determined spectroscopic parameters, a rotational temperature of 5~K, and a linewidth of 500~MHz.

\begin{figure*}
    \centering
    \includegraphics[width=\linewidth]{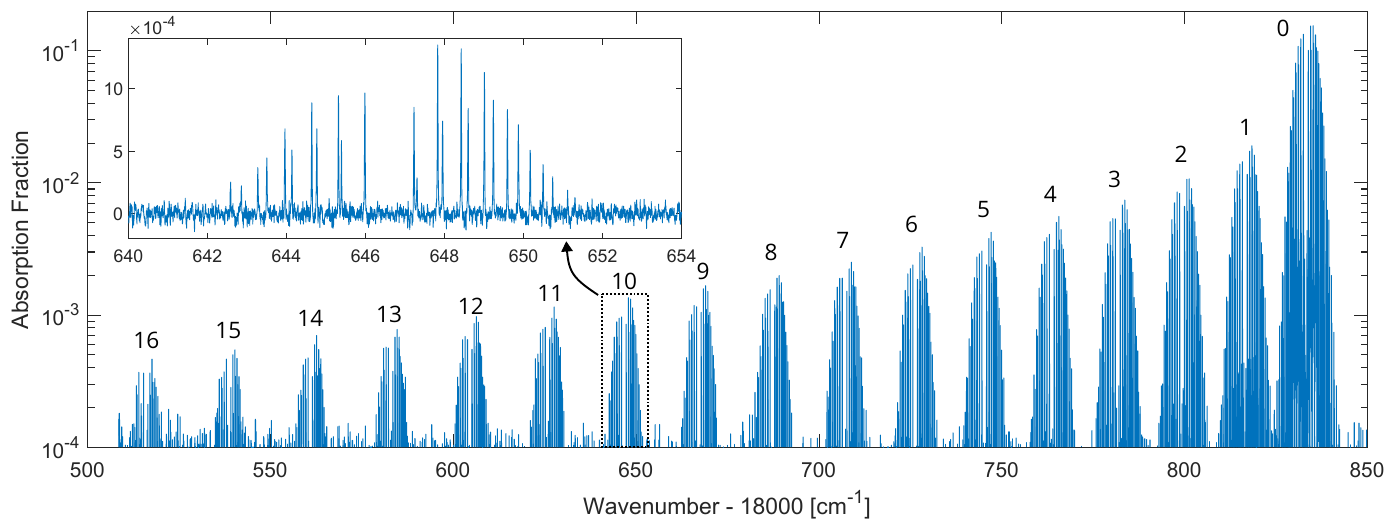}
    \caption{A broad absorption spectrum in the region of  $\Delta v=0$ bands of the $B^2\Sigma^+-X^2\Sigma^+$ electronic transition. The numbers above the features indicate the vibrational level of the ground and excited states.  The inset shows the $(10,10)$ band.  The data required around three hours to take.}
    \label{fig:CaFSpectrumWide}
\end{figure*}

\begin{figure}
    \centering
    \includegraphics[width=\linewidth]{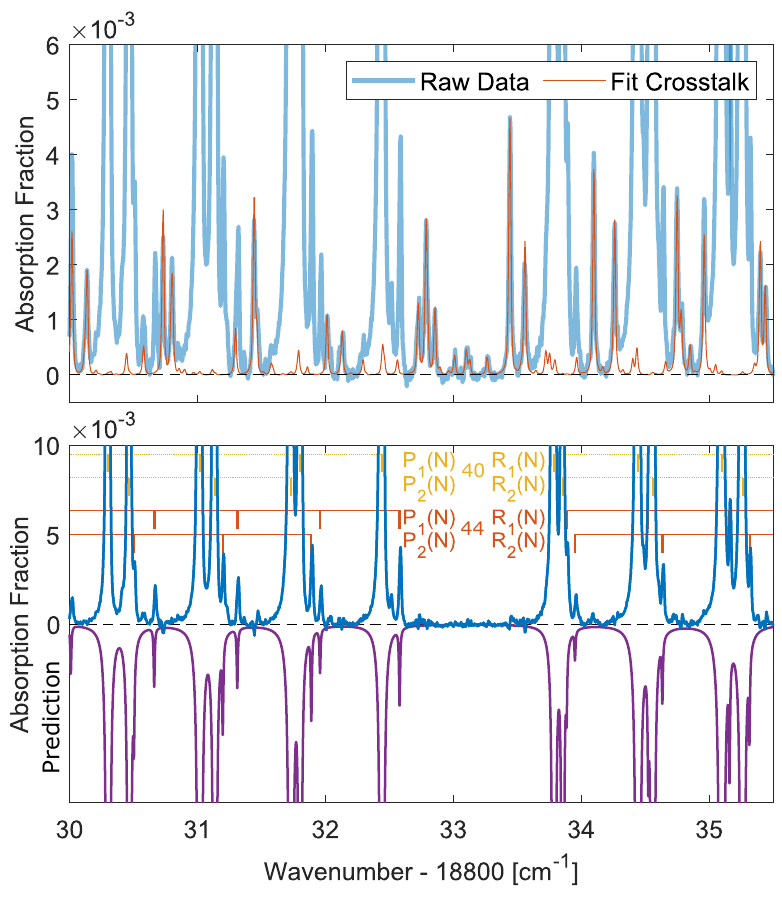}
    \caption{\textit{Top}: Modeling the crosstalk by fitting shifted versions of the spectrum.  \textit{Bottom}: The spectrum after subtracting fitted crosstalk with identification of $^{40}$CaF and $^{44}$CaF spectral features.  The prediction shown below the data has the features scaled by the natural abundance ratio.}
    \label{fig:CaFIsotopologues}
\end{figure}

A broad absorption spectrum in the region of $\Delta v=0$ bands of the $B^2\Sigma^+-X^2\Sigma^+$ electronic transition is given in Fig.~\ref{fig:CaFSpectrumWide}.  The absorption spectrum $(0,0)B^2\Sigma^+-X^2\Sigma^+$ band is presented in the bottom panel of Fig.~\ref{fig:CaFSpectrumWide} and a zoomed-in spectrum with assignments for both the $^{40}$CaF and $^{44}$CaF spectral features are given in Fig.~\ref{fig:CaFIsotopologues}.  The assignment of $^{44}$CaF lines is consistent with the expected --0.1176~cm$^{-1}$ shift of the origin, $T_{00}$ , and the approximately 3$\%$
reduction of the rotational constants, $B’$ and $B''$, obtained using standard mass scaling~\cite{Bernath2005Book}.  The list of parameters can be found in Table~\ref{tab:iso}.

Note that in order to properly interpret the spectrum in this region we must account for crosstalk between the VIPA orders, which will result in spurious duplication of strong features at multiples of the VIPA free spectral range (FSR) but with suppressed magnitude.  Our VIPA\footnote{LightMachinery OP-6721-3371-2, 3.37~mm thick} has a specified FSR of 1.00(1)~cm$^{-1}$, and the manufacturer reported a measured crosstalk in the first order FSR of $\sim 1.5\%$.  We model the crosstalk in the spectrum by fitting shifted versions of the spectrum peaks suppressed by $2.20\%,0.93\%$ for the $\pm$1 orders and $0.21\%,0.30\%$ for the $\pm$2 orders, respectively, where the FSR is 0.997~cm$^{-1}$ and all parameters were determined by fitting.

A perusal of the determined rotation, $B$, and spin-rotation, $\gamma$, parameters (see Table~\ref{tab:const}) extracted from analysis of the VIPA spectra shows that there is excellent agreement with those extracted from the combined analysis of newly recorded hollow-cathode emission spectrum and previously recorded LIF and pure-rotational spectra~\cite{Lavy2025}.

The sensitivity over a large dynamic range can be gauged from a comparison of the predicted and observed relative intensities of the $^{40}$CaF and $^{44}$CaF spectral features. The ratio of intensities is consistent with the relative abundance $^{40}$Ca (96.94\%) and $^{44}$Ca (2.086\%).

\begin{figure}
    \centering
    \includegraphics[width=\linewidth]{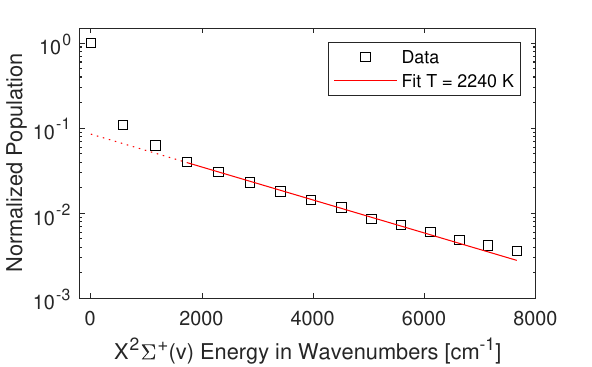}
    \caption{A plot of the $P_1(1)$ line intensities of the $^{40}$CaF isotopologue for $(0,0)B^2\Sigma^+-X^2\Sigma^+$ through $(14,14)B^2\Sigma^+-X^2\Sigma^+$ (see Table~\ref{tab:intensities}) as a function $X^2\Sigma^+$ vibrational energy. An exponential fit of the intensities for $v>3$ gives $T_\mathrm{vib}$=2,240~K.}
    \label{fig:vibPopulation}
\end{figure}

Unlike translation and rotation, the vibrational excitation is not effectively cooled in the CBGC source~\cite{Hutzler2012}. A list of the vibrational populations is shown in Table~\ref{tab:intensities}, and a plot is shown in Fig.~\ref{fig:vibPopulation}.  The higher levels are reasonably described by a thermal population with temperature around 2,000~K, though the lower levels have considerably higher population.  The ability to measure this wide range of both rotational and vibrational line strengths simultaneously opens up opportunities for internal state thermometry.

\section{\label{sec:applications} Outlook and Future Improvements}

There are a number of straightforward means to improve the operation of this spectrometer.  A cryogenic cell with larger windows could enable multi-pass configurations to increase the absorption fraction.  This would enable spectroscopy with even weaker lines, including weak transitions and less-abundant isotopes, which could be disentangled using laser-driven chemical enhancement methods~\cite{Pilgram2021YbOHOdd}.  The spectrometer resolution could be further increased by using a cavity to convert the white light source into a series of narrower peaks spaced by the spectrometer resolution which could then be scanned, similar to the approach used with frequency combs~\cite{Adler2010,Nugent-Glandorf2012VIPA}.

A number of molecules of interest have transitions in the visible and near-IR, and extending the range of the spectrometer over this range is critical for widespread use of this technology. In order to extend operation across the visible and near-IR range, a spectrometer with tunable range is in development at LightMachinery.  Such an expanded range would also make the spectrometer interesting for material characterization in addition to gas phase molecular spectroscopy.

In addition to absorption spectroscopy, the spectrometer could be used for dispersed laser-induced fluorescence.  The single-mode fiber input would require careful alignment and reduce collection efficiency, but the low dark count rate of the camera would allow for long integration times and be compatible with both cryogenic buffer gas beam and supersonic sources.  The high spectral resolution would enable direct measurement of rotational branching ratios, which are especially interesting for non-linear molecules~\cite{Augenbraun2020ATM}.

\begin{acknowledgments}

This work was supported by a DeLogi Science and Technology Grant.  We acknowledge assistance from John Hunter, Taylor Breen, and John Reid in understanding important technical details of the LightMachinery optical spectrometer.  Thanks to Chandler Conn, John M. Doyle, Zack Lasner, Ian Miller, John Reid, and Amar Vutha for giving feedback on the manuscript, and to Nachiket Bhanushali for assistance setting up the optical spectrometer.  We acknowledge helpful conversations with Mitchio Okumura and Termeh Bashiri.

\end{acknowledgments}

\section*{Data Availability Statement}

Some aspects of this work are the subject of a provisional patent application.  Data and analysis code are available for non-commercial use only.

\nocite{*}
\bibliography{references}

\onecolumngrid

\appendix

\newpage

\section{Lines and Intensities}

\begin{table*}[h]
\begin{tabular}{c|ll|ll|ll|ll}
 \toprule
Band  & \multicolumn{2}{c|}{$X^2\Sigma^+$} & \multicolumn{4}{c|}{$B^2\Sigma^+$}  \\
& \multicolumn{2}{c|}{$B$} & \multicolumn{2}{c|}{$B$} & \multicolumn{2}{c|}{$\gamma$} & ${T_{vv}}^a$ & $\sigma^b$ \\
$(v'', v')$ & Measured$^c$ & Previous & Measured$^c$ & Previous & Measured$^c$ & Previous \\ \hline
0,0   & 0.342483(37)  & 0.342488116(52)  & 0.341300(33)  & 0.3412892(19)  & --0.04569(13) & --0.045767(74) & 833.12699(51)  & 0.0011 \\
1,1   & 0.340040(49)  & 0.340053826(53)  & 0.338674(44)  & 0.3386754(21)  & --0.04617(17) & --0.045942(81) & 816.42847(68)  & 0.0015 \\
2,2   & 0.337655(42)  & 0.337629458(70)   & 0.336107(37)  & 0.3360826(14)  & --0.04641(15) & --0.046606(57) & 799.27456(58)  & 0.0012 \\
3,3   & 0.335214(45)  & 0.335215314(80)  & 0.333462(40)  & 0.3334921(43)  & --0.04687(16) & --0.04701(25)  & 781.67409(62)  & 0.0014 \\
4,4   & 0.332872(66)  & 0.332811563(112) & 0.330953(54)  & 0.33089528(49) & --0.04718(20) & --0.04728      & 763.47412(73)  & 0.0017 \\
5,5   & 0.330404(51)  & 0.330418358(181)  & 0.328347(43)  & 0.3283367(11)  & --0.04731(17) & --0.04767      & 745.15912(64)  & 0.0014 \\
6,6   & 0.327984(39)  & 0.32803600(23)   & 0.325763(31)  & 0.3257748(11)  & --0.04759(11) & --0.04806      & 726.26114(44)  & 0.0010 \\
7,7   & 0.325700(47)  & 0.32566419(35)   & 0.323270(45)  & 0.3232295(12)  & --0.04825(18) & --0.04846      & 706.94725(67)  & 0.0015 \\
8,8   & 0.323334(58)  & 0.32330425(51)    & 0.320728(52)  & 0.3206916(45)  & --0.04821(21) & --0.04874(18)  & 687.22414(76)  & 0.0016 \\
9,9   & 0.320875(61)  & 0.3209530       & 0.318160(53)  & 0.318167(90)   & --0.04861(21) & --0.04905(14)  & 667.10071(70)  & 0.0015 \\
10,10 & 0.318602(80)  & 0.3186130        & 0.315630(74)  & 0.31565611(63) & --0.04902(29) & --0.04963      & 646.58796(98)  & 0.0020 \\
11,11 & 0.316286(74)  & 0.3162836        & 0.313162(59)  & 0.31315992(66) & --0.04971(22) & --0.05003      & 625.69046(72)  & 0.0014 \\
12,12 & 0.314057(86)  & 0.3139648        & 0.310743(67)  & 0.3106552(67)  & --0.04996(25) & --0.05035(30)  & 604.41984(85)  & 0.0016 \\
13,13 & 0.311641(78)  & 0.3116566        & 0.308200(58)  & 0.3081795(36)  & --0.05057(22) & --0.0509(21)   & 582.78374(66)  & 0.0013 \\
14,14 & 0.309389(116) & 0.3093589        & 0.305697(123) & 0.3057075(87)  & --0.05097(50) & --0.05128(28)  & 560.79224(148) & 0.0024\\ \toprule
\end{tabular}
\caption{Determined spectroscopic parameters in wavenumbers (cm$^{-1}$) for the  $X^2\Sigma^+$ and $B^2\Sigma^+$ states.  Numbers in parentheses are 1$\sigma$ statistical estimates; entries without error estimates were values obtained using equilibrium constants. (a) The origin minus 18000.0~cm$^{-1}$. (b) Standard deviation of the fit. (c) The spin-rotation parameter, $\gamma$, for the $X^2\Sigma^+$ state was constrained to 1.3138$\times$10$^{-3}$~cm$^{-1}$ in the fit~\cite{Lavy2025}.\\
}
\label{tab:const}
\end{table*}

\begin{table*}[h]
\begin{tabular}{c|ll|ll}
 \toprule
  & \multicolumn{2}{c|}{$^{40}$CaF} & \multicolumn{2}{c}{$^{44}$CaF}  \\ 
Parameter~[cm$^{-1}$] & $X^2\Sigma^+(v=0)$ & $B^2\Sigma^+(v=0)$ & $X^2\Sigma^+(v=0)$ & $B^2\Sigma^+(v=0)$ \\ \hline
$B$	& 0.342483(37) & 0.341300(33) & 0.332460 & 0.331312 \\
$\gamma$ & 0.001313 (Fixed) & --0.04569(13) & 0.001272 & --0.044357 \\
$T$ & \; -- & 18833.12699(51) & \; -- & 18833.24459  \\
\toprule
\end{tabular}
\caption{The spectroscopic parameters used to model the  $(0,0)B^2\Sigma^+-X^2\Sigma^+$ band.  The $^{40}$CaF parameters were derived from fitting the observed spectral features and those for  $^{44}$CaF from scaling those parameters using reduced mass ratios.
}
\label{tab:iso}
\end{table*}

\begin{table*}[h]
    \centering
    \begin{tabular}{cccc}
        \toprule
        $(v, v)$ & Intensity & Normalized$^a$ & Energy (cm$^{-1}$)$^b$ \\
        \hline
        0,0   & 0.13330 & 1.0000 & 0    \\
        1,1   & 0.01446 & 0.1085 & 583  \\
        2,2   & 0.00835 & 0.0626 & 1160 \\
        3,3   & 0.00534 & 0.0401 & 1731 \\
        4,4   & 0.00408 & 0.0306 & 2297 \\
        5,5   & 0.00306 & 0.0230 & 2857 \\
        6,6   & 0.00240 & 0.0180 & 3411 \\
        7,7   & 0.00191 & 0.0143 & 3961 \\
        8,8   & 0.00156 & 0.0117 & 4505 \\
        9,9   & 0.00115 & 0.0086 & 5043 \\
        10,10 & 0.00097 & 0.0073 & 5576 \\
        11,11 & 0.00081 & 0.0061 & 6104 \\
        12,12 & 0.00065 & 0.0049 & 6626 \\
        13,13 & 0.00056 & 0.0042 & 7143 \\
        14,14 & 0.00048 & 0.0036 & 7655 \\
        \hline\\
    \end{tabular}
    \caption{Intensities of the $P_1(1)$ line of the $(v,v)B^2\Sigma^+-X^2\Sigma^+$ bands and $X^2\Sigma^+(v)$ energy.  (a) Normalized to the (0,0) intensity.  (b) The $X^2\Sigma^+(v)$ energy relative to $X^2\Sigma^+(v=0)$}
    \label{tab:intensities}
\end{table*}

\begin{table*}[h]
\footnotesize
\begin{tabular}{llllllllll}
Band & $N$ & 0 & 1 & 2 & 3 & 4 & 5 & 6 & 7 \\\hline
(0,0)$^a$
& $P_1(N)^b$ & \qquad --             & 832.4421(07)$^d$ & 831.7321(15)  & 831.0182(09)  & 830.3005(--13) & 829.5829(--10) & 828.8653(17)  & 828.1400(--09)    \\
& $P_2(N)$  & \qquad --             & \qquad --             & 831.8023(--01) & 831.1352(--09) & 830.4682(06)  & 829.7974(07)  & 829.1227(--07) & 828.4480(03)     \\
& $R_1(N)$  & 833.7842(--25) & 834.4436(02)  & 835.0992(14)  & 835.7509(11)  & 836.3988(--07) & 837.0467(--01) & 837.6907(--09) & 838.3348(06)     \\
& $R_2(N)^c$ & 833.8544(--08) & 834.5607(10)  & 835.2592(--18) & 835.9617(16)  & 836.6564(--03) & 837.3512(02)  & 838.0421(--08) & 838.7330(06)     \\
\hline (1,1) 
& $P_1(N)$  & \qquad --             & 815.7477(00)  & 815.0429(17)  & 814.3303(--16) & 813.6178(--21) & 812.9053(02)  & 812.1890(14)  & 811.4688(14)     \\
& $P_2(N)$  & \qquad --             & \qquad --             & 815.1130(--07) & 814.4510(37)  & 813.7852(--22) & 813.1194(--07) & 812.4498(--03) & 811.7763(--10)    \\
& $R_1(N)$  & 817.0836(09)  & 817.7341(05)  & 818.3807(--10) & 819.0274(02)  & 819.6702(04)  & 820.3091(--06) & 820.9482(12)  & 821.5794(--20)    \\
& $R_2(N)$  & 817.1537(17)  & 817.8509(--01) & 818.5443(--23) & 819.2377(--18) & 819.9312(15)  & 820.6169(--01) & 821.3027(09)  & 821.9847(09)     \\
\hline (2,2) 
& $P_1(N)$  & \qquad --             & 798.6010(24)  & 797.8974(11)  & 797.1900(--10) & 796.4826(02)  & 795.7715(06)  & 795.0564(02)  & 794.3376(--09)    \\
& $P_2(N)$  & \qquad --             & \qquad --             & 797.9712(20)  & 797.3105(--11) & 796.6498(--11) & 795.9891(21)  & 795.3207(06)  & 794.6484(--16)    \\
& $R_1(N)$  & 799.9227(--08) & 800.5681(--07) & 801.2097(--13) & 801.8513(12)  & 802.4851(--08) & 803.1190(02)  & 803.7491(05)  & 804.3753(01)     \\
& $R_2(N)$  & 799.9927(--05) & 800.6848(--21) & 801.3769(02)  & 802.0652(17)  & 802.7457(--15) & 803.4263(--15) & 804.1069(17)  & 804.7799(02)     \\
\hline (3,3)  
& $P_1(N)$  & \qquad --             & 781.0043(13)  & 780.3059(09)  & 779.6037(02)  & 778.8977(--08) & 778.1917(18)  & 777.4780(02)  & 776.7605(--18)    \\
& $P_2(N)$  & \qquad --             & \qquad --             & 780.3796(10)  & 779.7240(--13) & 779.0683(00)  & 778.4089(09)  & 777.7418(--24) & 777.0785(17)     \\
& $R_1(N)$  & 782.3158(--18) & 782.9561(--08) & 783.5926(--02) & 784.2252(02)  & 784.8540(01)  & 785.4790(--02) & 786.1040(30)  & 786.7173   (--19) \\
& $R_2(N)$  & 782.3895(16)  & 783.0764(04)  & 783.7595(--06) & 784.4387(--19) & 785.1179(04)  & 785.7895(--15) & 786.4611(01)  & 787.1289(14)     \\
\hline (4,4)  
& $P_1(N)$  & \qquad --             & 762.8086(09)  & 762.1113(--26) & 761.4160(--02) & 760.7148(01)  & 760.0098(04)  & 759.3008(06)  & \qquad --                \\
& $P_2(N)$  & \qquad --             & \qquad --             & 762.1855(--24) & 761.5391(03)  & 760.8848(--10) & 760.2305(15)  & 759.5683(01)  & \qquad --                \\
& $R_1(N)$  & 764.1152(28)  & 764.7461(--02) & 765.3769(07)  & 8766.0039(15) & 766.6250(03)  & 767.2441(10)  & 767.8594(16)  & 768.4648(--37)    \\
& $R_2(N)$  & 764.1836(04)  & 764.8652(--09) & 765.5430(--17) & 766.2168(--25) & 766.8926(25)  & 767.5586(15)  & 768.2207(05)  & 768.8789(--06)    \\
\hline (5,5)  
& $P_1(N)$  & \qquad --             & 744.4987(11)  & 743.8069(--15) & 743.1152(01)  & 742.4158(--19) & 741.7164(03)  & 741.0094(--10) & \qquad --                \\
& $P_2(N)$  & \qquad --             & \qquad --             & 743.8803(--23) & 743.2388(09)  & 742.5896(05)  & 741.9366(05)  & 741.2798(07)  & 740.6191(12)     \\
& $R_1(N)$  & 745.7936(14)  & 746.4198(--07) & 747.0461(15)  & 747.6646(00)  & 8748.2832(27) & 748.8942(18)  & 749.5013(12)  & 750.1007(--30)    \\
& $R_2(N)$  & 745.8631(00)  & 746.5396(--10) & 747.2123(--12) & 747.8811(--10) & 748.5462(--05) & 749.2074(03)  & 749.8648(13)  & 750.5145(--12)   \\
\hline (6,6)   
& $P_1(N)$   & \qquad --             & 725.6067(22)  & 724.9202(05)  & 724.2298(--05)   & 723.5357(--09) & 722.8377(--07) & 722.1360(02)  & \qquad --               \\
& $P_2(N)$   & \qquad --             & \qquad --             & 724.9935(--09) & 724.3533(--07)   & 723.7092(01)  & 723.0614(16)  & 722.4059(--02) & \qquad --               \\
& $R_1(N)$   & 726.8874(--15) & 727.5124(09)  & 728.1296(00)  & 728.7431(--03)   & 729.3528(00)  & 729.9586(10)  & 730.5567(--13) & 731.1549(09)    \\
& $R_2(N)$   & 726.9607(04)  & 727.6320(--05) & 728.2994(--01) & 728.9630(08)    & 729.6190(--14) & 730.2750(09)  & 730.9234(--01) & 731.5679(--05)   \\
\hline (7,7)   
& $P_1(N)$   & \qquad --             & 706.2947(--05) & 705.6134(--07) & 704.9284(01)    & 704.2395(20)  & 703.5430(11)  & 702.8427(12)  & 702.1347(--14)   \\
& $P_2(N)$   & \qquad --             & \qquad --             & 705.6904(06)  & 705.0515(--20)   & 704.4127(04)  & 703.7662(--01) & 703.1120(--33) & 702.4617(22)    \\
& $R_1(N)$   & 707.5689(--08) & 708.1887(21)  & 708.7970(--16) & 709.4054(--04)   & 710.0099(18)  & 710.6068(12)  & 711.1960(--22) & 711.7853(--07)   \\
& $R_2(N)$   & 707.6420(00)  & 708.3081(--11) & 708.9703(--05) & 709.6287(12)    & 710.2795(01)  & 710.9264(00)  & 711.5696(10)  &                 \\
\hline (8,8)   
& $P_1(N)$   & \qquad --             & 686.5769(01)  & 685.9009(08)  & 685.2173(--10)   & 684.5337(24)  & 683.8386(--03) & 683.1398(--17) & 682.4372(--16)   \\
& $P_2(N)$   & \qquad --             & \qquad --             & 685.9739(--19) & 685.3440(06)    & 684.7065(06)  & 684.0652(21)  & 683.4163(11)  & \qquad --               \\
& $R_1(N)$   & 687.8407(--08) & 688.4516(--08) & 689.0586(--06) & 689.6618(15)    & 690.2574(13)  & 690.8453(--14) & 691.4333(11)  & 692.0136(12)    \\
& $R_2(N)$   & 687.9137(--01) & 688.5745(--10) & 689.2315(02)  & 689.8808(--10)   & 690.5302(30)  & 691.1681(08)  & 691.7984(--39) & \qquad --               \\
\hline (9,9)   
& $P_1(N)$   & \qquad --             & 666.4599(16)  & 665.7853(--08) & 665.1070(--16)   & 664.4249(--07) & 663.7390(18)  & 663.0416(--16) & \qquad --               \\
& $P_2(N)$   & \qquad --             & \qquad --             & 665.8620(--04) & 665.2373(26)    & 664.6012(--05) & 663.9612(--19) & 663.3214(22)  & \qquad --               \\
& $R_1(N)$   & 667.7133(06)  & 668.3190(03)  & 668.9209(17)  & 669.5151(09)    & 670.1018(--21) & 670.6884(04)  & 671.2675(06)  & \qquad --               \\
& $R_2(N)$   & 667.7823(--33) & 668.4417(--05) & 669.0934(08)  & 669.7375(00)    & 670.3778(07)  & 671.0105(06)  & 671.6394(--05) & \qquad --               \\
\hline (10,10) 
& $P_1(N)$   & \qquad --             & 645.9501(00)  & 645.2809(--09) & 644.6079(03)    & 643.9272(--01) & 643.2390(--22) & 642.5547(56)  & \qquad --               \\
& $P_2(N)$   & \qquad --             & \qquad --             & 645.3574(--12) & 644.7340(--06)   & 644.1031(--17) & 643.4684(--06) & 642.8261(--11) & \qquad --               \\
& $R_1(N)$   & 647.1931(--16) & 647.7936(--12) & 648.3904(13)  & 648.9757(--16)   & 649.5610(14)  & 650.1349(--11) & 650.7050(--14) & \qquad --               \\
& $R_2(N)$   & 647.2696(14)  & 647.9199(05)  & 648.5664(24)  & 649.2052   (27) & 649.8365(13)  & 650.4602(--17) & \qquad --             & \qquad --               \\
\hline (11,11)
& $P_1(N)$   & \qquad --             & 625.0580(08)  & 624.3941(12)  & 623.7225(02)    & 623.0434(--20) & 622.3644(20)  & \qquad --             & \qquad --               \\
& $P_2(N)$   & \qquad --             & \qquad --             & 624.4704(--04) & 623.8484(--27)   & 623.2266(12)  & 622.5933(00)  & \qquad --             & \qquad --               \\
& $R_1(N)$   & 626.2906(--14) & 626.8859(--06) & 627.4737(--11) & 628.0577(08)    & 628.6341(14)  & 629.2029(06)  & 629.7641(--15) & \qquad --               \\
& $R_2(N)$   & 626.3669(04)  & 627.0119(--08) & 627.6531(10)  & 628.2867(16)    & 628.9128(08)  & 629.5312(--14) & \qquad --             & \qquad --             \\
\hline (12,12) 
& $P_1(N)$   & \qquad --             & 603.7900(--11) & 603.1314(07)  & 602.4652(16)  & 601.7877(--24) & 601.1102(04)  & \qquad --             & \qquad --   \\
& $P_2(N)$   & \qquad --             & \qquad --             & 603.2075(--14) & 602.5947(15)  & 601.9704(--04) & 601.3424(05)  & \qquad --             & \qquad --   \\
& $R_1(N)$   & 605.0160(--04) & 605.6062(06)  & 606.1888(07)  & 606.7638(--03) & 607.3313(--22) & 607.8950(--12) & 608.4549(26)  & \qquad --   \\
& $R_2(N)$   & \qquad --             & 605.7319(--06) & 606.3678(15)  & 606.9962(26)  & 607.6131(--10) & 608.2264(--18) & \qquad --             & \qquad --   \\
\hline (13,13) 
& $P_1(N)$   & \qquad --             & 582.1604(05)  & 581.5033(--04) & 580.8424(17)  & 580.1703(--06) & 579.4944(03)  & \qquad --             & \qquad --   \\
& $P_2(N)$   & \qquad --             & \qquad --             & 8581.5830(02) & 580.9716(--02) & 580.3525(--12) & \qquad --             & \qquad --             & \qquad --   \\
& $R_1(N)$   & 583.3759(10)  & 583.9571(--13) & 584.5346(--06) & 585.1045(--05) & 585.6668(--11) & 586.2254(15)  & 586.7726(--05) & \qquad --   \\
& $R_2(N)$   & \qquad --             & 584.0863(--06) & 584.7131(--23) & 585.3400(30)  & 585.9518(--01) & 586.5598(00)  &               & \qquad --   \\
\hline (14,14)
& $P_1(N)$   & \qquad --             & 560.1755(26)  & 559.5199(--06) & 558.8569(--41) & 558.1977(37)  & 557.5196(--01) & \qquad --             & \qquad --   \\
& $P_2(N)$   & \qquad --             &               & 559.6033(30)  & 558.9933(03)  & 558.3757(--26) & \qquad --             & \qquad --             & \qquad --   \\
& $R_1(N)$   & 561.3767(--14) & 561.9566(05)  & 562.5251(--15) & 563.0899(01)  & 563.6471(15)  & 564.1930(--10) & \qquad --             & \qquad --   \\
& $R_2(N)$   & \qquad --             & 562.0817(--38) & 562.7108(26)  & 563.3249(12)  & 563.9314(--04) & \qquad --             & \qquad --             & \qquad --   \\
\end{tabular}
\caption{Observed and predicted transition wavenumbers for the $(v'=0-14,v''=0-14) B^2\Sigma^+-X^2\Sigma^+$ bands of CaF.
(a) Band assignment $(v',v'')$.
(b) The weaker satellite $PQ_{12}(N)$ lines are not resolved from the more intense $P_1(N)$ lines.
(c) The weaker satellite $RQ_{21}(N)$ lines are not resolved from the more intense $R_2(N)$ lines.
(d) Transition wavenumber (cm$^{-1}$) minus 18000 and the difference between observed and predicted wavenumbers obtained using optimized parameters.  For example, ``818.5443(--25)'' of the $R_2(2)$ of the (0,0) band was measured to be at 18818.5443~cm$^{-1}$ and predicted to be at 18818.5468~cm$^{-1}$.}
\label{tab:lines}
\end{table*}

\end{document}